\documentclass[12pt]{article}
\usepackage[russian]{babel}
\usepackage{graphicx}
\newcommand{\be}{\begin{equation}}
\newcommand{\ee}{\end{equation}}
\newcommand{\ba}{\begin{eqnarray}}
\newcommand{\ea}{\end{eqnarray}}

\begin{document}
\begin{center}
{\bf {SUPERSYMMETRIC  DYNAMICS  AND ZETA-FUNCTIONS
}}
\end{center}
\begin{center}
 Nugzar Makhaldiani
 \footnote{e-mail address:~~mnv@jinr.ru}
\end{center}
\begin{center}
Joint Institute for Nuclear Research, Dubna\\
\end{center}
Boson, fermion, and super oscillators and  (statistical) mechanism of cosmological constant; finite approximation of the zeta-function and fermion factorization of the bosonic statistical sum considered.

PACS:
11.30.Pb 
03.65.Yz 

%

\begin{flushright}
 \vspace{.5cm}


 I always knew that sooner or later p -\\
  adic numbers will appear in Physics - \\
Andr$\acute{e}$ Weil.\\

\end{flushright}



Supermathematics unifies discrete and continual aspects of mathematics.
Boson oscillator hamiltonian is
\ba
H_b=\hbar \omega(b^+b+bb^+)/2=\hbar \omega(b^+b+a),\ a=1/2.
\ea
corresponding energy spectrum $E_{bn}$ and eigenfunctions $|n_b>$ are
\ba
H_b|n_b>=E_{bn}|n_b>,\ E_{bn}=\hbar \omega(n_b+a),\ n_b=0,1,2,...
\ea
Fermion oscillator hamiltonian, eigenvectors and energies are
\ba
&&H_f=\hbar \omega(f^+f+ff^+)/2=\hbar \omega(f^+f-a),\cr
&& H_f=|n_f>=E_{fn}|n_f>,\ E_{fn}=\hbar \omega(n_f-a),\ n_f=0,1.
\ea
For supersymmetric oscillator we have
\ba
&&H=H_b+H_f,\ H|n_b,n_f>=\hbar \omega(n_b+n_f)|n_b,n_f>,\cr
&& |n_b,n_f>=|n_b>|n_f>,\ E_{n_b,n_f}=\hbar \omega(n_b+n_f)
\ea
For background-vacuum $|0,0>,$ energy  $E_{0,0}=0.$ For higher energy states $|n-1,1>,\ |n,0>,\ E_{n-1,1}=E_{n,0}.$
Supersymmetry needs not only the same frequency for boson and fermion oscillators, but also that $2a=1.$

A minimal realization of the algebra of supersymmetry
\ba \label{QH}
&&\{Q,Q^+\}=H,\
\{Q,Q\}=\{Q^+,Q^+\}=0,
\ea
is given by a point particle dynamics in one dimension, \cite{Witten1981}
\ba \label{QQ+}
&& Q=f(-iP+W)/\sqrt{2},\
Q^+=f^+(iP+W)/\sqrt{2},\ P=-i\partial /\partial x
\ea
where
the superpotential $W(x)$ is any function
of x, and spinor operators $f$ and $f^+$ obey the anticommuting relations
\ba \label{phi}
&& \{f,f^+\}=1, \
f^2=(f^+)^2=0.
\ea

There is a following representation of operators
$f$, $f^+$ and $\sigma$ by
 Pauli spin matrices
\ba \label{Pauli}
f=\frac{\sigma_1-i\sigma_2}{2}=\left(
                                  \begin{array}{cc}
                                    0 & 0 \\
                                    1 & 0 \\
                                  \end{array}
                                \right),
f^+ =\frac{\sigma_1+i\sigma_2}{2}=\left(
                                      \begin{array}{cc}
                                        0 & 1 \\
                                        0 & 0 \\
                                      \end{array}
                                    \right),
\sigma=\sigma_3=\left(
                    \begin{array}{cc}
                      1 & 0 \\
                      0 & -1 \\
                    \end{array}
                  \right)
\ea

From formulae (\ref{QH}) and (\ref{QQ+}) then we have
\ba \label{H}
H=(P^2+W^2+\sigma W_{x})/2.
\ea

The simplest nontrivial case of the superpotential $W=\omega x$
corresponds to the supersimmetric oscillator with Hamiltonian \ba
H=H_b+H_f, \ \ H_b=(P^2+\omega^2x^2)/2,\ \ H_f=\omega\sigma/2,
\ea
The ground state energies of the bosonic
and fermionic parts are \ba E_{b0}=\omega/2,\ \ E_{f0}=-\omega/2, \ea
so the vacuum energy of the supersymmetric oscillator is
\ba
<0|H|0>=E_0=E_{b0}+E_{f0}=0,\ \ |0>=|n_b,n_f>=|n_b>|n_f>.
\ea


Let us see on this toy - solution of the cosmological constant problem from
the quantum statistical viewpoint. The statistical sum of the
supersymmetric oscillator is \ba Z(\beta)=Z_bZ_f, \ea where
\ba
&&Z_b=\sum_{n}^{}e^{-\beta E_{bn}}=e^{-\beta\omega/2}+e^{-\beta\omega(1+1/2)}+...=e^{-\beta\omega/2}/(1-e^{-\beta\omega})\cr
&&Z_f=\sum_{n}^{}e^{-\beta
E_{fn}}=e^{\beta \omega/2}+e^{-\beta\omega/2}. 
\ea
In the low
temperature limit, \ba Z(\beta)=1+O(e^{-\beta \omega})\rightarrow
1,\ \ \beta =T^{-1}, \ea
so cosmological constant
$\lambda\sim lnZ\rightarrow 0.$
From observable values of $\beta$ and the cosmological constant we estimate $\omega.$

The Riemann zeta function (RZF) can be interpreted in thermodynamic terms as a statistical sum of a system with energy spectrum: $E_n=\ln n,\ n=1,2,...:$
\ba
\zeta(s)=\sum_{n\geq1}n^{-s}=Z(\beta)=\sum_{n\geq1}\exp(-\beta E_n),\ \beta=s,\ E_n=\ln n,\ n=1,2,...
\ea


Let us consider the following finite approximation of RZF
\ba
&&\zeta_N(s)=\sum_{n=1}^Nn^{-s}=\frac{1}{\Gamma(s)}\int_0^\infty dtt^{s-1}\frac{e^{-t}-e^{-(N+1)t}}{1-e^{-t}}=\zeta(s)-\Delta_N(s),\ Re\ s>1\cr
&&\zeta(s)=\frac{1}{\Gamma(s)}\int_0^\infty dt\frac{t^{s-1}}{e^{t}-1},\ \Delta_N(s)=\frac{1}{\Gamma(s)}\int_0^\infty dt\frac{t^{s-1}e^{-Nt}}{e^{t}-1}
\ea

Another formula, which can be used on critical line, is
\ba
\zeta(s)&=&(1-2^{1-s})^{-1}\sum_{n\geq 1}(-1)^{n+1}n^{-s}\cr
&=&\frac{1}{1-2^{1-s}}\frac{1}{\Gamma(s)}\int_0^\infty\frac{t^{s-1}dt}{e^{t}+1},\ Re\ s>0
\ea
Corresponding finite  approximation of RZF is
\ba
&&\zeta_N(s)=(1-2^{1-s})^{-1}\sum_{n=1}^N(-1)^{n-1}n^{-s}\cr
&&=\frac{1}{1-2^{1-s}}\frac{1}{\Gamma(s)}\int_0^\infty\frac{t^{s-1}(1-(-e^{-t})^N)dt}{e^{t}+1}=\zeta(s)-\Delta_N(s),\cr
&& \Delta_N(s)=\frac{1}{\Gamma(s)}\int_0^\infty dt\frac{t^{s-1}(-e^{-t})^N)}{e^{t}+1}\sim\pm N^{-s},\cr
&&|s|\simeq\frac{|\ln \Delta_N(s)|}{\ln N}
\ea
at a (nontrivial) zero of RZF, $s_0,$ $\zeta_N(s_0)=-\Delta_N(s_0).$ In the integral form, dependence on $N$ is analytic and we can consider any complex valued $N.$ It is interesting to see dependence (evolution) of zeros with $N,$
\ba
\frac{d\Delta_N(s)}{dN}=(-1)^{N-1}s\Delta_N(s+1)
\ea

For the simplest nontrivial integer $N=2,$
\ba
&&\zeta_2(s)=(1-2^{1-s})^{-1}(1-2^{-s})\cr
&&=\frac{1-2^{-s}}{1-2^{1-s}}=\frac{2^s-1}{2^s-2}=\frac{2^{s-1/2}-1/\sqrt{2}}{2^{s-1/2}-\sqrt{2}},
\ea
we have
zeros at $s=2\pi in/\ln 2,\ n=0, \pm1, \pm2,...$


Let as consider the following formula (Qvelementar particles)
\ba\label{pbfs}
\frac{1}{1-q}=(1+q)(1+q^2)(1+q^4)...,\ |q|<1.
\ea
which can be proved as
\ba
&&p_k\equiv(1+q)(1+q^2)(1+q^4)...(1+q^{2^k})=\frac{1-q^{2^{(k+1)}}}{1-q},\cr
&&c(1-|q|^{2^{(k+1)}})<|p_k|<c(1+|q|^{2^{(k+1)}}),\ \lim_{k\rightarrow\infty}|p_k|=c=1/|1-q|,\cr
&& \lim_{k\rightarrow\infty}p_k=1/(1-q).
\ea
The formula (\ref{pbfs}) reminds us the boson and fermion statsums
\ba
Z_b=\frac{q^a}{1-q},\ Z_f=\frac{1+q}{q^a},\ q=\exp{(-\beta\omega)},\ a=1/2,\ \beta=1/T
\ea
and can be transformed in the following relation
\ba
Z_b(\omega)=Z_f(\omega)Z_f(2\omega)Z_f(4\omega)...
\ea
Indeed,
\ba
&&Z_b(\omega)=\frac{q^a}{1-q}=q^bZ_f(\omega)Z_f(2\omega)Z_f(4\omega)...,\cr
&& b=2a+2a(1+2+2^2+...)=2a(1+\frac{1}{1-2})=0,\ |2|_2=1/2,
\ea
where $|n|_p=1/p^k,\ n=p^km,$ is p-adic norm of $n,\ k$ is the number of $p$ - prime factors of $n$.

Bytheway we have an extra bonus! We see that the fermion content of the boson wears the p-adic sense \cite{MakhaldianiSQS2009}. The prime $p=2,$ in this case. Also, the vacuum energy of the oscillators wear p - adic sense.

We may consider also the following recurrent relations
\ba
&&Z_b(\omega)=\frac{q^a}{1-q}=\frac{1+q}{q^a}\frac{q^{2a}}{1-q^2}=Z_f(\omega)Z_b(2\omega)=Z_f(\omega)Z_f(2\omega)Z_b(2^2\omega)=\cr
&&...=Z_f(\omega)Z_f(2\omega)...Z_f(2^{k-1}\omega)Z_b(2^k\omega)
\ea
and ask that
\ba
&&Z_b(2^{k}\omega)=1\Rightarrow q_k^2+q_k-1=0\Rightarrow q_k=\exp(-2^k\beta\omega)=\frac{\sqrt{5}-1}{2}=\frac{2}{\sqrt{5}+1}
\cr
&&
T_k=\beta_k^{-1}=2^k\omega/\ln g,\ k=0,1,2,3,...,\
g=\frac{\sqrt{5}+1}{2}=1.618,\cr
&& \ln g=0.4812,\ 1/\ln g=2.078=2+\varepsilon,\ \varepsilon=0.078,\cr
&& T_k=(\omega_{k+1}+\varepsilon\omega_k),\ \omega_k=2^k\omega
\ea
so, at the temperatures $T=T_k,$ we have the finite  representation of the boson statistical sum by fermion statistical sums. For higher values of $k$ and corresponding higher temperatures there are more fermion factors. We may also consider an invariant combination:
$T_k/\omega=2^k/\ln g.$
Note that $Z_b(\omega_0)=Z_f(\omega_0),\ T_0=\omega/\ln g.$ As we have seen, to $Z_b=1$ corresponds zero cosmological constant.
It is curious to identify $T_0$ with relict radiation temperature.

Let us consider the simplest extension of one level fermion system,
\ba
Z_{2f}=q^\alpha(1+q+q^2),\ q=\exp(-\beta\omega)
\ea
Now we find the vacuum energy-the parameter $\alpha,$ from the following construction
\ba
&&1+q+q^2=\frac{1-q^3}{1-q}\Rightarrow q^\alpha(1+q+q^2)\frac{q^{3/2}}{1-q^3}=\frac{q^{1/2}}{1-q}\Downarrow\cr
&& Z_{2f}(\omega)Z_b(3\omega)=Z_b(\omega),\ \alpha=-1
\ea
Corresponding p-state generalization is
\ba
Z_{pf}(\omega)Z_b((p+1)\omega)=Z_b(\omega),\ Z_{pf}(\omega)=q^{-p/2}(1+q+...+q^p)
\ea

For RZF we have the following factorized representation by prime numbers 
\ba
\zeta_b(s)=\sum_{n\geq1}n^{-s}
=\prod_p(1-p^{-s})^{-1}=(\prod_k p_k)^{-as}\prod_pZ_{bp}(s),\ Z_{bp}(s)=\frac{p^{as}}{1-p^{-s}}
\ea
it hints on boson statsums.
For this we consider a fermion zeta function
\ba
&&\zeta_f(s)=\sum_{n=p_1^{n_1}...p_k^{n_k}}n^{-s}
=\prod_p(1+p^{-s})=(\prod_k p_k)^{as}\prod_pZ_{fp}(s),\ n_k=0, 1\cr
&& Z_{fp}(s)=\frac{1+p^{-s}}{p^{as}}
\ea
Now we have
\ba
Z(s)=Z_fZ_b=\zeta_f(s)\zeta_b(s)=\prod_p\frac{1+p^{-s}}{1-p^{-s}}
\ea
We may consider also the following parafermion extensions
\ba
&&\zeta_f(s)=\sum_{m=p_1^{n_1}...p_k^{n_k}}m^{-s}
=\prod_p(1+p^{-s}+...+p^{-s(n-1)})=(\prod_k p_k)^{as}\prod_pZ_{fpn}(s),\cr
&& Z_{fpn}(s)=\frac{1+p^{-s}+...+p^{-s(n-1)}}{p^{as}},\ n_k=0, 1,...,n-1
\ea
\ba
Z(s)=Z_fZ_b=\zeta_{fpn}(s)\zeta_b(s)=\prod_p\frac{1+p^{-s}+...+p^{-s(n-1)}}{1-p^{-s}}
\ea

A braid group $B_{N+1}$ is generated by elements $s_n, n = 1, . . . N,$ subject to
relations:
\ba
s_ns_{n+1}s_n=s_{n+1}s_ns_{n+1}, \ s_ns_m=s_ms_n,\ m\neq n\pm 1
\ea
An A-Type Hecke algebra $H_{N+1} = H_{N+1}(q)$ (see, e.g., \cite{Wenzl} and references
therein) is a quotient of the group algebra of the braid group $B_{N+1}$ by a Hecke
relations
\ba
s_n^2=1+\lambda s_n,\ n=1,...N,\ \lambda=q-q^{-1},
\ea
where $q\in C/\{0\}$ is a parameter. Note that, for $\lambda=-1,$
\ba
&&q^2+q-1=0,\cr
&&s_n^2+s_n-1=0.
\ea

As another example where similar structure appears, let us consider, on the base of familiar sequence, the following discrete dynamical system
\ba\label{Fib}
f_{n+1}=f_n+f_{n-1}
\ea  
which reduce when $f_n\neq0$ to the following one
\ba
\varphi_{n+1}=1+1/\varphi_n
,\ \varphi_n=f_n/f_{n-1}
\ea 
with convergent points $x_{\pm}:$
\ba
&&x_{+}=1+\frac{1}{1+x_{+}}=1+\frac{1}{1+\frac{1}{1+x_{+}}}=...=1+\frac{1}{1+\frac{1}{1+...}}=\frac{\sqrt{5}-1}{2},\cr
&&x_{-}
=-1-x_{+}=-1+\frac{1}{x_{-}}=-1+\frac{1}{-1+\frac{1}{-1+...}}=-\frac{\sqrt{5}+1}{2}
\ea
We may consider this dynamical system as exactly solvable example of one charge renormdynamics with ultraviolet and infrared fixed points as $x_{-}$ and  $x_{+}.$

It is not obvious that the dynamical system (\ref{Fib}) is reversible: knowing $f_{n+1}$ we can not define $f_n.$ On the level of pares $(f_n,f_{n-1})$
we obtain explicitly reversible system:
\ba
F_{n+1}=AF_{n-1},\ F_n=\left(
                       \begin{array}{c}
                         f_n \\
                         f_{n-1} \\
                       \end{array}
                     \right),\ A=\left(
                                   \begin{array}{cc}
                                     2 & 1 \\
                                     1 & 1 \\
                                   \end{array}
                                 \right),\ detA=1.
\ea 
So, the formulation with explicit time reflection symmetry needs one, the simplest, step of renormdynamics: double time steps. On this level appears internal degrees of freedom-spin.





\end{document}